\documentclass[fleqn,usenatbib,useAMS]{mnras}

\usepackage{mathtools}

\usepackage{breqn}
\usepackage[T1]{fontenc}
\usepackage{ae,aecompl}
\usepackage{tablefootnote}

\usepackage{graphicx}	
\usepackage{amsmath}	
\usepackage{amssymb}	
\usepackage{subfig}
\usepackage{float}
\usepackage{multirow}
\usepackage{units}
\usepackage{placeins}

\usepackage{xcolor}

\title[Spectro-polarimetric view of LMC X$-3$]{First detection of X-ray polarization in thermal state of LMC X-3: Spectro-polarimetric study with {\it IXPE}}

\author[Majumder et al.]{Seshadri Majumder$^{1}$\thanks{E-mail: smajumder@iitg.ac.in}, Ankur Kushwaha$^{2}$\thanks{E-mail: ankurksh@ursc.gov.in},
	Santabrata Das$^{1}$\thanks{E-mail: sbdas@iitg.ac.in  (Corresponding Author)}, 
	Anuj Nandi$^{2}$\thanks{E-mail: anuj@ursc.gov.in}\\
	$^{1}$Department of Physics, Indian Institute of Technology Guwahati, Guwahati, 781039, India.\\
	$^{2}$Space Astronomy Group, ISITE Campus, U. R. Rao Satellite Centre, Outer
	Ring Road, Marathahalli, Bangalore, 560037, India.
}

\date{Accepted XXX. Received YYY; in original form ZZZ}

\pubyear{2023}

\begin{document}
	\label{firstpage}
	\pagerange{\pageref{firstpage}--\pageref{lastpage}}
	\maketitle

\begin{abstract}
	
We report a comprehensive spectro-polarimetric study of the black hole binary LMC X$-3$ using simultaneous {\it IXPE}, {\it NICER} and {\it NuSTAR} observations in 0.5$-$20 keV energy band. The broad-band energy spectrum (0.5$-$20 keV) with {\it NICER} and {\it NuSTAR} is well described by the disc emission of temperature $\sim$1.1 keV and a weak Comptonizing tail beyond $\sim$10 keV. This evidently suggests a disc-dominated spectral state of the source with disc contribution of $\sim$96\%. The lack of variability ($rms$$\sim$0.5\%) in the power spectrum further corroborates the high/soft nature of the source. A significant polarization degree (PD) of 3.04$\pm$0.40\% (>7$\sigma$) at a polarization angle (PA) of $-$44.24$^{\circ}$$\pm$3.77$^{\circ}$ (>7$\sigma$) is found in 2$-$8 keV energy range of {\it IXPE}. In addition, PD is seen to increase with energy up to $\sim$4.35$\pm$0.98\% (>3$\sigma$) in 4$-$8 keV band. Further, we attempt to constrain the source spin ($a_{*}$) using broad-band spectral modelling that indicates a weakly rotating `hole' in LMC X$-3$ with $a_{*}=0.273_{-0.012}^{+0.011}-0.295_{-0.021}^{+0.008}$ ($90\%$ confidence). Based on the spectro-polarimetric results, we infer that the polarization in LMC X$-3$ is resulted possibly due to the combined effects of the direct and/or reflected emissions from a partially ionized disc atmosphere. Finally, we discuss the relevance of our findings.
	
\end{abstract}



	
\section{Introduction}

The {\it Imaging X-ray Polarimetry Explorer} \cite[{\it IXPE};][]{Weisskopf-etal2022} opens up a new window for understanding the emission processes and geometry of the accreting objects in a more sophisticated way by investigating the X-ray polarimetric properties since its launch in $2021$. The high quality observations that is capable of measuring the degree of polarization along with the net direction of the electric field vector in the polarized emission challenges the current understandings of the environment around the black hole X-ray binaries (BH-XRBs).

Meanwhile, the detection of polarized emission in a few BH-XRBs including Cyg X$-1$ \cite[]{Krawczynski-etal2022,Dovciak-etal2023,Jana-etal2023}, 4U 1630$-$47 \cite[]{Ratheesh-etal2023, Kushwaha-etal2023, Rawat-etal2023} and Cyg X-3 \cite[]{Veledina-etal2023} observed with {\it IXPE} are confirmed in their different spectral states. In particular, for Cyg X$-1$, a high polarization degree (PD) of $\lesssim 20\%$ ($130-230$ keV) is observed with {\it INTEGRAL/SPI} \cite[]{Jourdain-etal2012}  as well as $\sim 75\%$ ($>400$ keV) observed with {\it INTEGRAL/IBIS} \cite[]{Rodriguez-etal2015}, whereas PD is found to be $< 10\%$ ($19-181$ keV) in the hard state when observed with {\it PoGo}+ \citep{Chauvin-etal2019}. Recently, \cite{Chattopadhyay-etal2023} reported a polarization fraction of $\sim 23\%$ ($100-380$ keV) in the hard intermediate state of Cyg X-1 with {\it AstroSat/CZTI}. However, the signature of polarized emission is not seen from LMC X$-1$ \cite[]{Podgorny-etal2023}, the only extra-galactic BH-XRBs observed by {\it IXPE}. Note that, the polarized emission from a geometrically thin and optically thick Novikov-Throne disc \cite[]{Novikov-etal1973} is better understood assuming Chandrasekhar's prescription for the scattering of light in semi-infinite atmospheres \cite[]{Podgorny-etal2023}. Moreover, the energy dependent polarimetric properties are also reported very recently that evidently indicates the presence of more complex physical processes yielding such scenario \cite[]{Kushwaha-etal2023, Jana-etal2023}. Indeed, the polarization properties are used to comprehend the geometry of the X-ray corona, an enduring problem till date \cite[]{Krawczynski-etal2022}.

Considering all these, for the first time to the best of our knowledge, we intend to examine the polarization properties of a persistent extra-galactic BH-XRBs named LMC X$-3$ using {\it IXPE} observations. The source is discovered by UHURU in $1971$ and located in Large Magellanic Cloud (LMC) at a distance of $48.1$ kpc \cite[]{Orosz-etal2009}. It harbours a slowly rotating black hole of mass $6.98\pm0.56 ~ {\rm M}_\odot$ \cite[]{Orosz-etal2014, Bhuvana-etal2021} and spin $0.25_{-0.29}^{+0.20}$ \cite[]{Steiner-etal2010, Bhuvana-etal2021} with a high disc inclination of $69.24^{\circ}\pm0.72^{\circ}$ \cite[]{Orosz-etal2014}. LMC X$-3$ mostly remains in high/soft state \cite[]{Bhuvana-etal2021} except for few occasions during which the low/hard states are observed \cite[]{Smale-etal2012} followed by an anomalous low state with luminosity $\sim 10^{35}$ erg $\rm s^{-1}$ \cite[]{Torpin-etal2017}. Hence, LMC X$-3$ remains a favourable candidate to study the polarization properties with {\it IXPE} due to its dominant soft spectral nature in low energy X-ray band.

Accordingly, in this Letter, we report the polarization results of LMC X$-3$ in thermally-dominated state observed with {\it IXPE} in $2-8$ keV energy range. While carrying out the spectro-polarimetric study of LMC X$-3$, we make use of the simultaneous broad-band coverage of {\it NICER} and {\it NuSTAR} in $0.5-20$ keV and constrain the black hole spin parameter.

The Letter is organized as follows: In \S2, we mention the observation details and data analysis procedures for each instrument. In \S3, we present the results obtained from spectro-polarimetric studies with {\it IXPE}, {\it NICER} and {\it NuSTAR}. Finally, we conclude with discussion in \S 4.

\section{Observation and Data Reduction}

LMC X$-3$ is observed during $7-20$ July $2023$ by {\it IXPE} \cite[]{Weisskopf-etal2022}, an X-ray polarimeter consisting of three identical detector units (DUs) operating in $2-8$ keV energy range for a total exposure of about $562$ ks. We analyze level-2 cleaned event files obtained from each DUs using \texttt{IXPEOBSSIMv30.0.0} software \cite[]{Baldini-etal2022}. Following the standard procedures \cite[]{Chatterjee-etal2023,Kushwaha-etal2023, Jana-etal2023}, we extract the event lists from a $60^{\prime \prime}$ circular source region centered at the source position and an annular background region between $180^{\prime \prime}$ and $240^{\prime \prime}$ radii of same center using the task \texttt{XPSELECT} in software routine. We use the tool \texttt{XPBIN} with algorithm \texttt{PCUBE} for model-independent polarimetric analysis. Similarly, I, Q and U spectra of {\it IXPE} are obtained with the algorithms \texttt{PHA1}, \texttt{PHA1Q} and \texttt{PHA1U}, respectively. The latest response files (v12) provided by the software team is used during fitting of Stokes spectra. The {\it IXPE} light curves of each DUs are extracted in \texttt{XSELECT} from respective level-2 clean event files.

{\it NICER} monitoring of the source is carried out on 8 July, 2023, one day after the start of {\it IXPE} observation. We process {\it NICER} data using \texttt{NICERDASv10} available in \texttt{HEASOFT V6.31.1}\footnote{\url{https://heasarc.gsfc.nasa.gov/docs/software/heasoft/}}. The cleaned event files are generated considering the standard data screening criteria with \texttt{nicerl2} which are further used to generate the spectral products using the task \texttt{nicerl3-spect} with background model \texttt{3c50}.

{\it NuSTAR} observed LMC X$-3$ six times during July, 2023 which overlap with that of {\it IXPE} observation. However, at the time of writing this letter, only one observation  on 9 July, 2023 having duration of $\sim 28$ ks was publicly available. We analyze the {\it NuSTAR} data using the dedicated software \texttt{NUSTARDAS}\footnote{\url{https://heasarc.gsfc.nasa.gov/docs/nustar/analysis/}} integrated in \texttt{HEASOFT V6.31.1}. We use the task {\it nupipeline} to generate the cleaned event files from both {\it FPMA} and {\it FPMB} instrument onboard {\it NuSTAR}. A circular region of 50 arcsec radii at the source position and away from it are considered to extract the source and background spectra, instrument response and ancillary files using the task {\it nuproduct}, respectively. All the {\it NICER} and {\it NuSTAR} spectra are grouped with 25 counts per bin to have a good statistics in the spectral fitting.

\begin{figure}
	\begin{center}
		\includegraphics[width=\columnwidth]{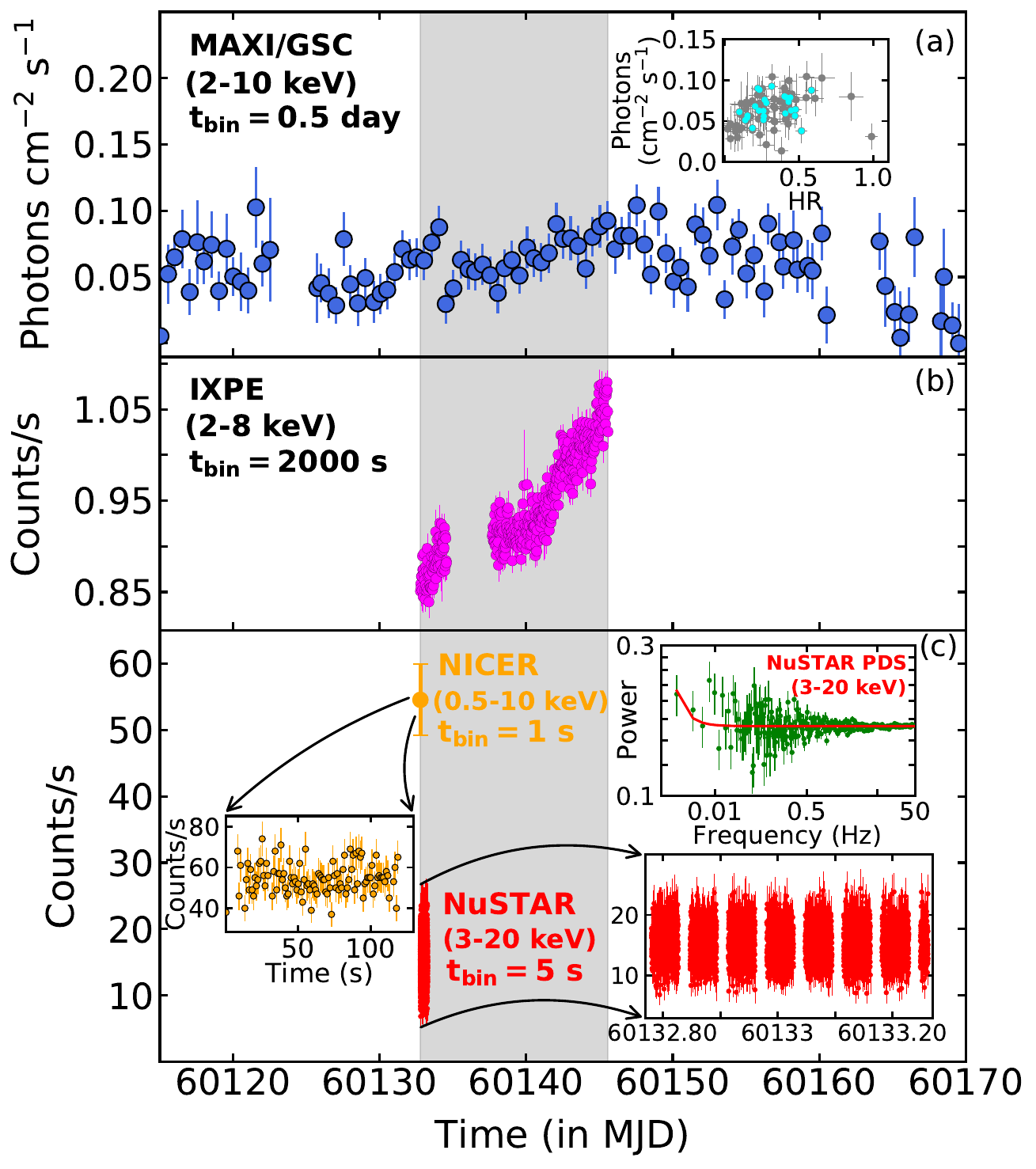}
	\end{center}
	\caption{Long-term monitoring of LMC X$-3$ obtained from different instruments in different energy bands. In panel (a), the {\it MAXI/GSC} lightcurve in $2-10$ keV energy band of $0.5$ day time bin along with the hardness intensity diagram is shown in the top right inset. Panel (b) depicts $2000$ s binned {\it IXPE} average light curve in $2-8$ keV energy band combining all the DUs. In panel (c), {\it NICER} (1 s bin) and {\it NuSTAR} ($5$ s bin) lightcurves are presented in $0.5-10$ keV and $3-20$ keV energy bands, respectively. The PDS obtained from {\it NuSTAR} data is presented in the top right inset of panel (c). See the text for details.}
	\label{fig:max_lc}
\end{figure}

\begin{figure}
	\begin{center}
		\includegraphics[width=\columnwidth]{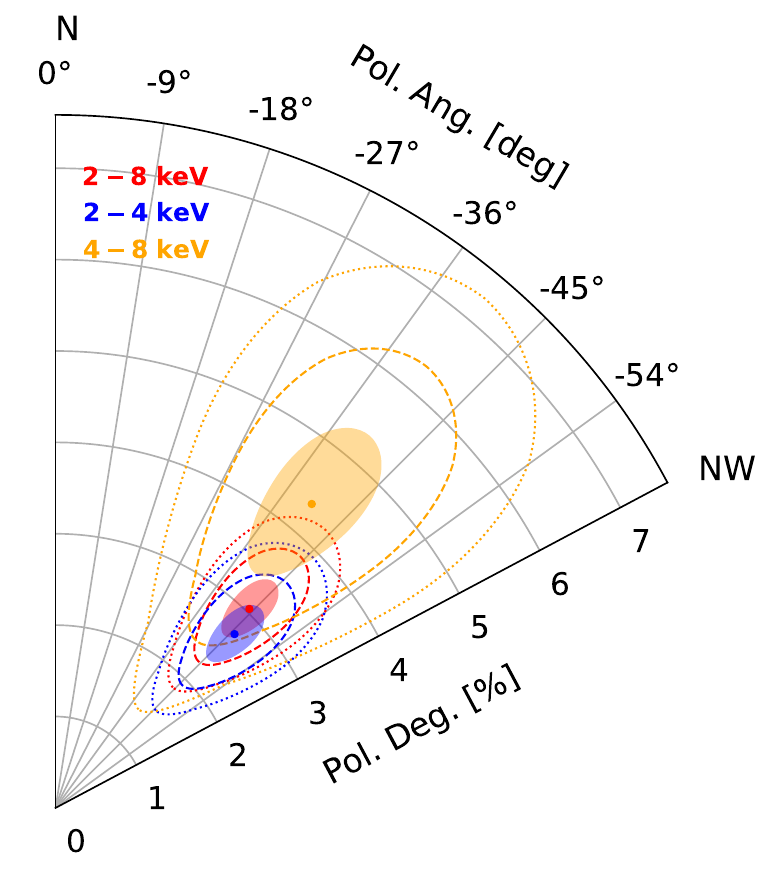}
	\end{center}
	\caption{Confidence contours of energy dependent polarization parameters of LMC X$-3$ obtained with model-independent analysis. Shaded region and regions bounded with dashed and dotted curves represent $1\sigma$, $2\sigma$ and $3\sigma$ confidence contours. Different energy ranges are marked in different colors. See the text for details.}
	\label{fig:contour_plot}
\end{figure}

\section{Analysis and Results}

\subsection{Variability Properties}

We study the long-term variability of the source in respective energy bands of different instruments. While doing so, we consider $0.5$ day averaged {\it MAXI/GSC} light curve in $2-10$ keV energy band (see Fig. \ref{fig:max_lc}a). The hardness-intensity diagram (HID) with hardness ratio ($HR$) (ratio of count rates in $4-10$ keV to $2-4$ keV energy bands) is presented at the top right inset. We observe that during {\it IXPE} observation the source persistently remains bright with a {\it MAXI/GSC} mean count rate of $\sim 0.07$ photons $\rm cm^{-2}$ $\rm s^{-1}$. In addition, the {\it MAXI}-HID during the {\it IXPE} observation (cyan circles) clearly indicates a high/soft nature of the source with a $HR \lesssim 0.5$. Further, we investigate 2000 s binned light curve of {\it IXPE} combining all the DUs in $2-8$ keV energy range (see Fig. \ref{fig:max_lc}b). Interestingly, the variability is observed with a fractional variability ($F_{\rm var}$) \citep{Vaughan-etal2003} of $\sim 4.84\%$ and $4.69\%$ in {\it IXPE} and {\it MAXI} lightcurves, respectively. However, negligible variability is seen with $F_{\rm var}\sim 1.98\%$ and $2.58\%$ in {\it NICER} and {\it NuSTAR} light curves due to less exposer. Further, we make use of $3-20$ keV {\it NuSTAR} data to obtain the power density spectrum (PDS) in $0.001-50$ Hz frequency range using \texttt{MaLTPyNT}\footnote{\url{https://www.matteobachetti.it/maltpynt/}}. The PDS shows no significant variability with rms amplitude of $0.5\%$ in $0.001-50$ Hz frequency range. This further confirms the high/soft nature of the source during the multi-mission campaign \cite[]{Remillard-Mcclintock2006}.

\subsection{Polarimetric Properties}

\begin{table}
	\centering
	\caption{Results from model-independent polarimetric study of LMC X$-3$ in different energy bands. Here, PD, PA, Q/I, U/I, MDP and SIGNIF denote polarization degree, angle of polarization, normalized Q-Stokes parameter, normalized U-Stokes parameter, minimum detectable polarization, and detection significance, respectively. $F_{\rm disc}$ ($\times 10^{-10}$ erg $\rm cm^{-2}$ $\rm s^{-1}$) refers disc flux obtained from broad-band energy spectral modelling.}
	\begin{tabular}{lccc}
		\hline
		
		Parameters &  $2-4$ keV & $4-8$ keV &$2-8$ keV  \\
		
		& (I)  & (II) &  \\
		
		\hline 
		
		PD(\%)  & $2.73 \pm 0.40$ & $4.35 \pm 0.98$ & $3.04 \pm 0.40$ \\
		
		PA ($^\circ$)  & $-45.85 \pm 4.19$ & $-40.13 \pm 6.43$ & $-44.24 \pm 3.77$\\
		
		Q$/$I (\%)  & $-0.08 \pm 0.40$ & $0.74 \pm 0.98$ & $0.08 \pm 0.40$ \\
		
		U$/$I (\%)  & $-2.73 \pm 0.40$ & $-4.29 \pm 0.98$ & $-3.04 \pm 0.40$ \\
		
		MDP (\%)  & $1.21$ & $2.96$ & $1.21$ \\
		
		SIGNIF ($\sigma$) & $6.41$ & $3.90$ & $7.20$\\ \hline
		
		$F_{\rm disc}$  $^\boxtimes$& $2.35$ & $1.10$ & $3.46$  \\	\hline
		
	\end{tabular}
	\begin{list}{}{}
		\item $^\boxtimes$ Calculated from the broad-band energy spectra in $0.5-20$ keV energy range fitted with \texttt{const*Tbabs*(diskbb$+$nthcomp)}.
	\end{list}
	
	\label{tab:en-pol}
\end{table}

We investigate the polarization properties of LMC X$-3$ for the first time using the {\it IXPE} observation conducted for a duration of $\sim 562$ ks. We carry out model-independent polarimetric analysis combining all events from the three detector units (DUs) of {\it IXPE} using \texttt{PCUBE} algorithm and estimate the polarization angle (PA), polarization degree (PD) and minimum detectable polarization (MDP). We find significant detection of polarization in $2-8$ keV energy band with PD$=3.04 \pm 0.40\%$ ($> 7 \sigma$) at MDP$=1.21\%$ alongwith a PA $=-44.24^{\circ} \pm 3.77^{\circ}$ ($>7\sigma$), measured from north to west in the sky. We further examine the energy dependent polarization properties in different energy bands, namely (I) $2-4$ keV and (II) $4-8$ keV, respectively. The obtained results are tabulated in Table \ref{tab:en-pol}. We observe that PD increases with energy and reaches to $4.35 \pm 0.98\%$ ($>3\sigma$) in $4-8$ keV energy band. We present the polarimetric results obtained from the different energy bands in Fig. \ref{fig:contour_plot}.

\begin{figure}
	\begin{center}
		\includegraphics[width=\columnwidth]{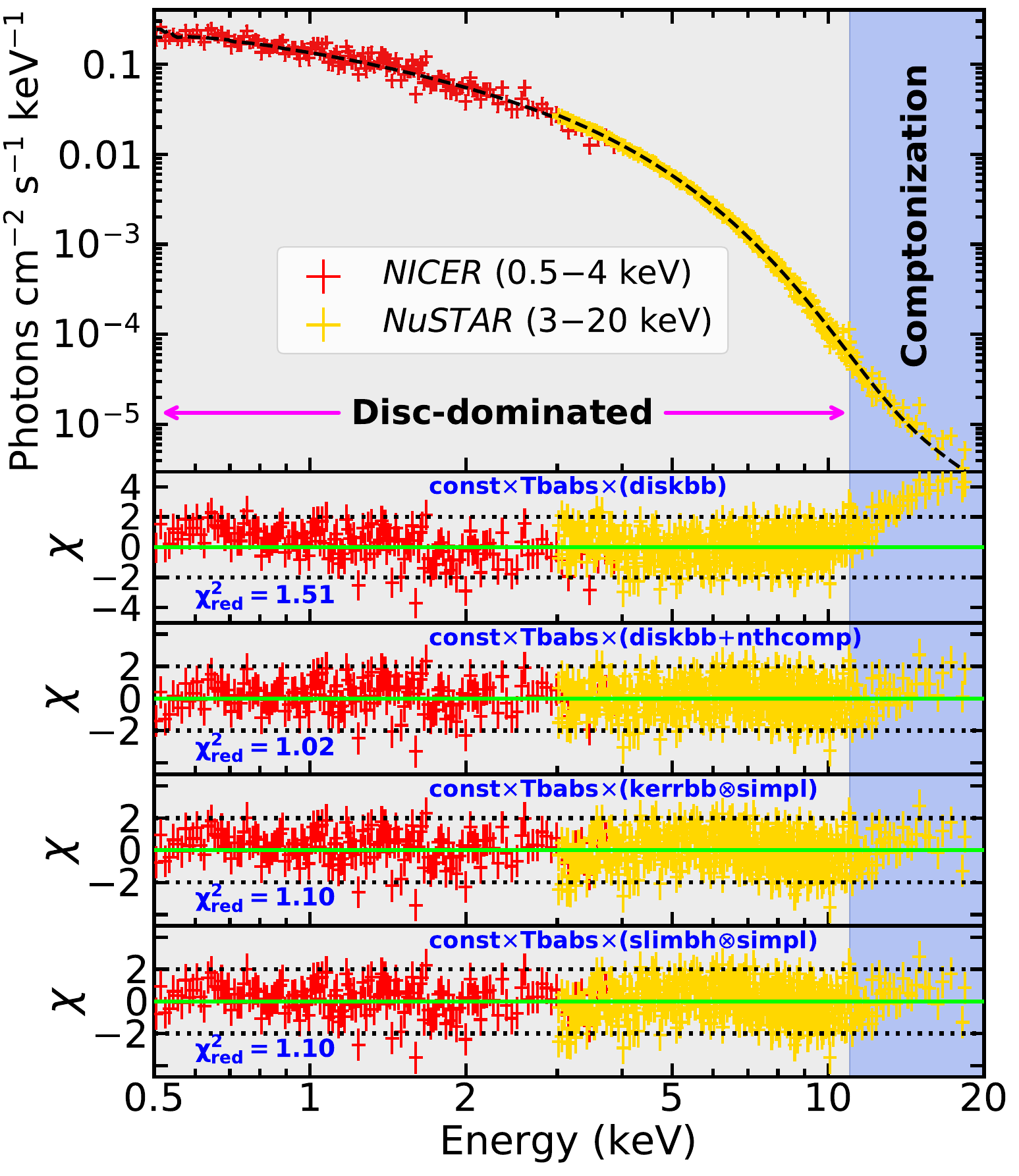}
	\end{center}
	\caption{{\it Upper panel:} Best fitted broad-band energy spectrum obtained from the simultaneous {\it NICER} ($0.5-4.0$ keV) and {\it NuSTAR} ($3-20$ keV) observations. {\it Lower panels:} Variation of the fitted residuals obtained from four different model prescriptions. }
	\label{fig:spec_nicer}
\end{figure}

\subsection{Broad-band Spectral Modelling}

\begin{table}
	\centering
	\caption{Best fitted spectral parameters obtained from the modelling of combined {\it NICER} and {\it NuSTAR} broad-band ($0.5-20$ keV) energy spectra with various model prescriptions (M$1$:\texttt{const*Tbabs*(diskbb$+$nthcomp)}; M$2$:\texttt{const*Tbabs*(kerrbb$\otimes$simpl)}; M$3$: \texttt{const*Tbabs*(slimbh$\otimes$simpl)}). Model components and parameters in the table have their usual meanings. All the errors are computed with 90\% confidence level. See the text for details.}
	
	\renewcommand{\arraystretch}{1.2}
	
	\resizebox{1.0\columnwidth}{!}{%
		\begin{tabular}{l @{\hspace{0.25cm}} l @{\hspace{0.1cm}} c @{\hspace{0.1cm}} c @{\hspace{0.2cm}} c } 
			\hline
			
			Model & Model & \multicolumn{3}{|c|}{Different model prescriptions} \\
			\cline{3-5}
			
			Components & Parameters & M1 & M2 & M3 \\	\hline 
			
			\texttt{Tbabs} & $n_{\rm H} (\times 10^{22})$ $\rm cm^{-2}$ & $0.04^{\dagger}$ & $0.04^{\dagger}$ & $0.04^{\dagger}$ \\	\hline
			
			 \texttt{diskbb} & $kT_{\rm in}$ (keV) & $1.09_{-0.01}^{+0.01}$ & $-$ & $-$ \\			
			&  $N_{\rm diskbb}$ &  $23.11_{-1.04}^{+1.06}$ &  $-$ &  $-$ \\			
			&  $F_{\rm disc} (\times 10^{-10})$ $\rm erg$ $\rm cm^{-2}$ $\rm s^{-1}$ &  $6.81_{-0.02}^{+0.02}$ &  $-$ & $-$ \\	\hline
			
			\texttt{nthcomp} & $kT_{\rm e}$ (keV) &  $10^{*}$ &  $-$ &  $-$ \\			
			& $\Gamma_{\rm nth}$ & $2.97_{-0.13}^{+0.10}$ & $-$ & $-$ \\			
			& $F_{\rm nth} (\times 10^{-10})$ $\rm erg$ $\rm cm^{-2}$ $\rm s^{-1}$  & $0.37_{-0.01}^{+0.01}$ & $-$ & $-$ \\ 		
			& $F_{\rm test}$ Probability $(\times 10^{-49})$ & $1.83$  & $-$ & $-$ \\	\hline
			
			\texttt{kerrbb} & $a_{*}$ & $-$ & $0.295_{-0.021}^{+0.008}$ & $-$  \\			
			& $M_{\rm BH}$ ($M_\odot$) & $-$ & $6.98^{\dagger}$ & $-$ \\			
			& $i$ (degree) & $-$ & $69.24^{\dagger}$ & $-$  \\			
			& $\dot{M} (\times 10^{18})$ (g $\rm s^{-1}$) & $-$ & $4.27_{-0.13}^{+0.12}$ & $-$ \\			
			& $d$ (kpc) & $-$ & $48.1^{\dagger}$ & $-$ \\  			
			& $f$ & $-$ & $1.7^{\dagger}$  & $-$ \\ \hline
			
			\texttt{slimbh} & $a_{*}$ & $-$ & $-$ & $0.273_{-0.012}^{+0.011}$  \\			
			& $M_{\rm BH}$ ($M_\odot$) & $-$ & $-$ & $6.98^{\dagger}$ \\			
			& $i$ (degree) & $-$ & $-$ & $69.24^{\dagger}$  \\			
			& $d$ (kpc) & $-$ & $-$ & $48.1^{\dagger}$ \\  			
			& $f$ & $-$ & $-$  & $1.7^{\dagger}$ \\		
			& $L_{\rm disc}$ ($L_{\rm Edd}$) & $-$ & $-$ & $0.27\pm0.01$ \\		
			& $\alpha$ & $-$ & $-$ & $0.1^{\dagger}$  \\ \hline
			
			\texttt{simpl} & $\Gamma_{\rm simpl}$ & $-$ & $1.09_{-0.02}^{+0.01}$ & $1.09_{-0.01}^{+0.01}$ \\
			& $f_{\rm scatt}$ ($\%$) & $-$ & $0.83_{-0.46}^{+0.11}$ & $0.84_{-0.45}^{+0.11}$ \\
			& $F_{\rm test}$ Probability $(\times 10^{-26})$ & $-$ & $1.04$ & $0.11$ \\ \hline
			
			& $\chi^2$/d.o.f & $571/564$ & $620/565$ & $621/564$ \\
			& $\chi^2_{\rm red}$ & $1.01$ & $1.10$ & $1.10$ \\	\hline
		\end{tabular}  
	} 	
	\begin{list}{}{}
		\item $^{\dagger}$Frozen to fiducial values.
		\item $^{*}$Not constrained within $1\sigma$ error; frozen at best fitted values. 
	\end{list}	
	\label{tab:spectral3}	
\end{table}

We study the broad-band spectral properties of the source using quasi-simultaneous observations with {\it NICER} and {\it NuSTAR} in $0.5-20$ keV energy band. While doing so, we adopt the standard disc model \texttt{diskbb} \cite[]{Makishima-etal1986} as the source of primary emissions and fit the spectra with the model \texttt{const*Tbabs*diskbb}. Here, \texttt{Tbabs} \cite[]{Wilms-etal2000} takes care of the galactic absorptions and \texttt{const} accounts for the cross calibration between different spectra. The obtained fit is resulted a reduced chi-square ($\chi^2_{\rm red}=\chi^2/d.o.f$) of $1.51$ with large residuals above $\sim 10$ keV confirming the presence of a high energy tail (see Fig. \ref{fig:spec_nicer}). 

Next, we consider thermal Comptonization model \texttt{nthcomp} \cite[]{Zdziarski-etal1996} in presence of a multi-temperature disc (as the seed photon distribution) to describe the high energy excess and find a good fit as $\chi^2_{\rm red}=1.02$. The inclusion of \texttt{nthcomp} component is found significant with $F$-test\footnote{\url{https://heasarc.gsfc.nasa.gov/xanadu/xspec/manual/node82.html}} probability of $1.83\times 10^{-49}$. We obtain the best fitted model parameters, such as inner disc temperature $kT_{\rm in} = 1.09_{-0.01}^{+0.01}$ keV and photon index $\Gamma_{\rm nth} = 2.97_{-0.13}^{+0.10}$. As the electron and seed photon temperatures remain unconstrained, we freeze these parameters at $10$ keV and $0.2$ keV, respectively. For the same reason, we keep the hydrogen column density fixed at $0.04\times10^{22}$ $\rm cm^{-2}$ \cite[]{Yilmaz-etal2023}. With this, we estimate the unabsorbed flux in $0.5-20$ keV energy range using \texttt{cflux} and find that $\sim 96\%$ of the total flux is associated with the disc component. Moreover, the higher disc flux contribution and a steeper photon index ($\sim 2.97$) evidently indicate that the source remains in the disc-dominated state. All the model fitted and estimated parameters obtained from the broad-band spectral fitting with model prescription \texttt{const*Tbabs*(diskbb+nthcomp)} are tabulated in Table. \ref{tab:spectral3}.

\begin{table*}
	\caption{Results from spectro-polarimetric analysis of the {\it IXPE} Stokes spectra with different models in $2-8$ keV energy range.}
		\begin{tabular}{l l c c c c c}
			\hline
			
			Components & $A_{\rm norm}$  & $A_{\rm index}$ & $\psi_{\rm norm}$ ($^\circ$) & $\psi_{\rm index}$ & $\rm PD$ ($\%$) & $\rm PA$ ($^\circ$) \\	\hline
			
			\texttt{polpow} & $0.011 \pm 0.001$ & $-0.82^\dagger$ &$-59.28 \pm 7.82$ & $0.24^\dagger$ & $4.08 \pm 0.37$ & $-41.11 \pm 5.43$ \\ \hline
			
			\texttt{polconst} & --- & --- & --- & --- & $2.94 \pm 0.61$ & $-44.96 \pm 6.01$\\
			
			\hline 
			
		\end{tabular}
	\begin{list}{}{}
		\item \hskip 2.3 cm $^\dagger$Frozen parameter as it could not be constrained within $1\sigma$ error.
	\end{list}
	\label{tab:pol_para}
\end{table*}

\subsection{Estimation of Spin}

The broad-band energy spectra of LMC X$-3$ reveals its high/soft nature, and hence, we attempt to estimate the spin of the source using continuum fitting method in $0.5-20$ keV energy range. In doing so, we rely on the relativistic accretion disc model \texttt{kerrbb} \cite[]{Li-etal2005} and following \cite{Orosz-etal2009, Orosz-etal2014}, we consider source distance $d=48.1$ kpc, mass  $M_{\rm BH} =6.98 ~ {\rm M}_\odot$ and inclination $i=69.24^{\circ}$ during the fitting. We fix the color factor ($f$) to a standard value of $1.7$. With this, the spin of the black hole ($a_{*}$) and the mass accretion rate ($\dot{M}$) are left as free parameter. In addition, we add \texttt{simpl} model \cite[]{Steiner-etal2009} to account the  Comptonizing tail above $\sim 10$ keV. The model \texttt{const*Tbabs*(kerrbb$\otimes$simpl)} is found to provide an excellent fit with a $\chi^{2}_{\rm red}$ of $1.10$. The resultant fit yields the source spin as $a_{*}=0.295_{-0.021}^{+0.008}$ (at $90\%$ confidence) and mass accretion rate as $\dot{M} = 4.27_{-0.13}^{+0.12} \times 10^{18}$ g~s$^{-1}$. We also find the photon index of \texttt{simpl} as $1.09_{-0.02}^{+0.01}$. Further, we employ the slim disc model \texttt{slimbh} \cite[]{Sadowski-etal2011} to estimate the source spin, where viscosity parameter is set as $\alpha = 0.1$. The resultant fit renders $\chi_{\rm red}^2=1.10$ that provides $a_{*}=0.273_{-0.012}^{+0.011}$ (at $90\%$ confidence) and disc luminosity $L_{\rm disc} = 0.27 \pm 0.01 ~ {\rm L}_{\rm Edd}$. Note that in explaining the high energy tail of the spectrum, the addition of \texttt{simpl} component is found significant with $F$-test probability of $1.04\times 10^{-26}$. Interestingly, both models provide consistent spin values that suggest LMC X$-3$ seems to harbour a weakly rotating black hole at its central core. We present the broad-band spectra of LMC X$-3$ in $0.5-20$ keV energy range in the top panel of Fig. \ref{fig:spec_nicer} and the fitted residuals obtained from different model combinations are shown in the lower successive panels. All the model fitted and estimated parameters obtained from the broad-band spectral fitting with model prescriptions \texttt{const*Tbabs*(kerrbb$\otimes$simpl)} and \texttt{const*Tbabs*(slimbh$\otimes$simpl)} are shown in Table. \ref{tab:spectral3}.

\subsection{Spectro-polarimetric Results}

\begin{figure}
	\begin{center}
		\includegraphics[width=\columnwidth]{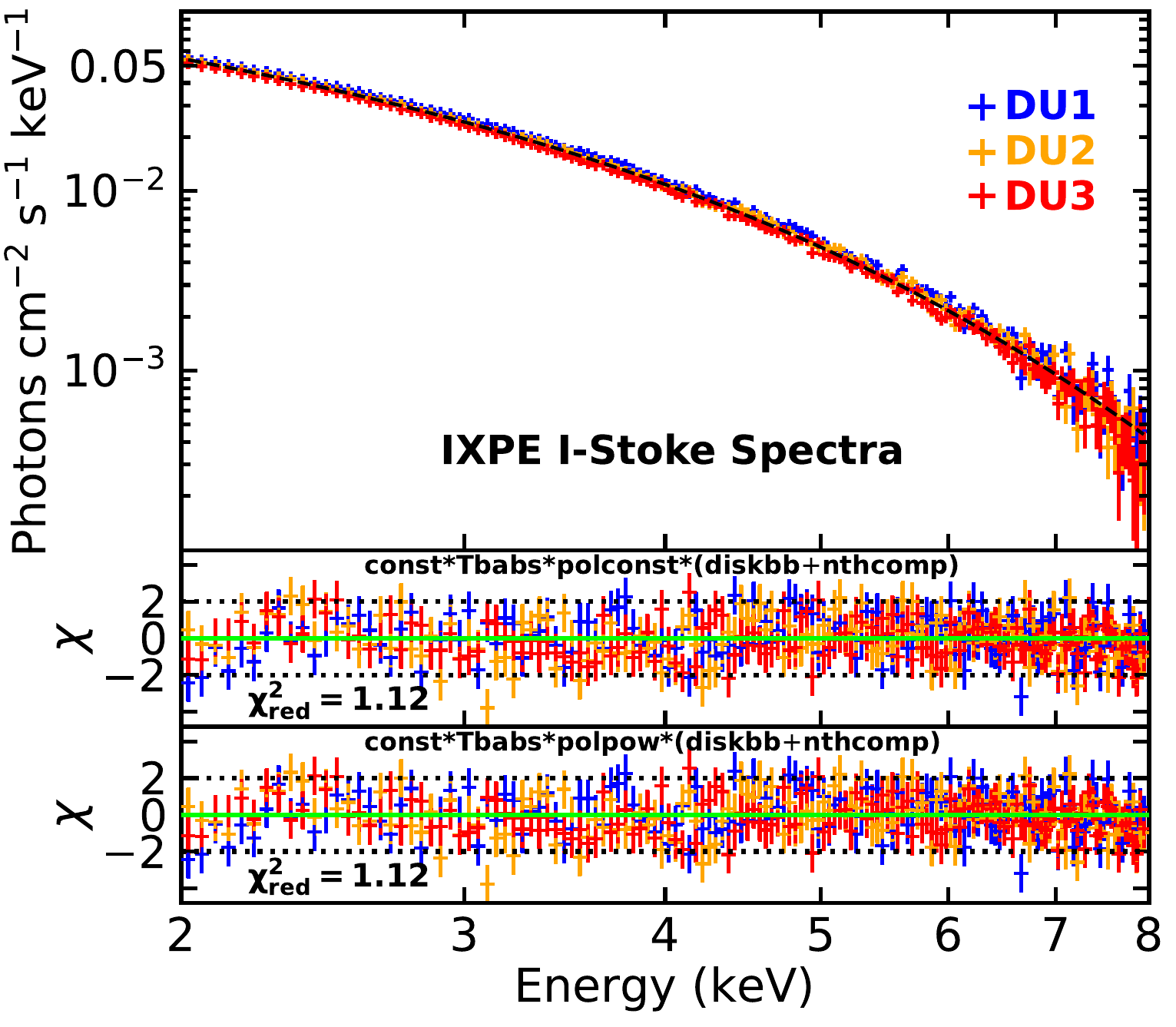}
	\end{center}
	\caption{{\it Top}: Best fit I-Stokes spectra of LMC X$-3$ obtained from model-dependent spectro-polarimetric study in $2-8$ keV energy band. {\it Middle-Bottom}: Residuals from different model combinations mentioned in both panels. Blue, orange and red denote results for DU1, DU2 and DU3, respectively.  See text for details.}
	\label{fig:spec_ixpe}
\end{figure}

We carry out the spectro-polarimetric analysis to examine the model-dependent polarization properties of LMC X$-3$. We simultaneously fit nine Stokes (I, Q and U) spectra obtained from all the three DUs of {\it IXPE} in $2-8$ keV energy range using \texttt{XSPEC}. As a first attempt, we adopt a model combination \texttt{const*Tbabs*polconst*(diskbb$+$nthcomp)} to fit the spectra. Here, \texttt{polconst} stands for the constant polarization model with PA and PD as the model parameters. The best fit with $\chi^2_{\rm red}=1.12$ yields ${\rm PD} = 2.94 \pm 0.61\%$ and ${\rm PA} = -44.96^{\circ} \pm 6.01^{\circ}$. As PA and PD are seen to vary with energy, we further consider energy dependent polarization model \texttt{polpow} to fit the spectra. The model uses PD($E$) $=A_{\rm norm}\times E^{-A_{\rm index}}$ (in fraction) and PA($E$) $=\psi_{\rm norm}\times E^{-\psi_{\rm index}}$ (in $^\circ$). We fit the Stokes spectra using the model combination \texttt{const*Tbabs*polpow*(diskbb$+$nthcomp)}. While fitting, $A_{\rm index}$ and $\psi_{\rm index}$ are fixed at their best fitted values of $-0.82$ and $0.24$, respectively as the parameters could not be constrained within $1\sigma$ errors. This yields best fit spectra with $\chi^{2}_{\rm red}=1.12$, $A_{\rm norm}=0.011 \pm 0.001$ and $\psi_{\rm norm} = -59.28^{\circ} \pm 7.82^{\circ}$, respectively. We integrate the model in $2-8$ keV energy range with the best fitted parameters and obtain PD $\sim 4.08 \pm 0.37\%$ and PA $\sim -41.11^{\circ} \pm 5.43^{\circ}$ which differs marginally with the results obtained from model-independent studies (see Table \ref{tab:en-pol}). All polarization parameters obtained from spectro-polarimetric study are tabulated in Table \ref{tab:pol_para}. The best fitted I-Stoke spectra with different models are shown in Fig. \ref{fig:spec_ixpe}.

\section{Discussion}

In this Letter, we report the results of a spectro-polarimetric study of LMC X$-3$ in broad energy range ($0.5-20$ keV) from the first ever {\it IXPE} observation along with the simultaneous observations of {\it NICER} and {\it NuSTAR}. The source is found to be in a thermally-dominated spectral state ($kT_{\rm in} \sim 1.09$ keV) with $\sim 96\%$ disc contribution and no significant variability ($0.5\%$) in the power spectrum.

The most intriguing results that we find in this study are the detection of significant polarized emission of ${\rm PD} = 3.04 \pm 0.40\%$ ($> 7 \sigma$) with ${\rm PA} = -44.24^{\circ} \pm 3.77^{\circ}$ ($ > 7 \sigma$). The spectro-polarimetric study from {\it IXPE} Stokes spectra yields a ${\rm PD} = 4.08 \pm 0.37\%$ and ${\rm PA} = -41.11^{\circ} \pm 5.43^{\circ}$ that marginally differs from the findings of model-independent studies (see Table \ref{tab:en-pol} and Table \ref{tab:pol_para}). These findings evidently suggest that LMC X$-3$ is the first ever detected extra-galactic BH-XRBs that exhibits polarization properties. The energy dependent polarimetric study indicates an increase of PD up to $\sim 4.35 \pm 0.98\%$ in $4-8$ keV energy band. Similar energy dependent polarization properties are also reported in several BH-XRBs including 4U 1630$-$47 \cite[]{Kushwaha-etal2023} and Cyg X$-1$ \cite[]{Jana-etal2023}. We observe negligible 
variation of PA over the different energy bands. We report the above findings for the first time for extra-galactic BH-XRBs.

It may be noted that the observed PD ($\sim 3.04\%$) of LMC X$-3$ is found higher compared to the same obtained in the soft state of Cyg X$-1$ ($\sim 2.79\%$) \cite[]{Jana-etal2023}. This possibly happens due to the fact that Cyg X$-1$ is a low inclination source than the LMC X$-3$ \cite[]{Schnittman-etal2009}. Meanwhile, highly polarized (PD $\sim 8.33\%$) emission in the disc-dominated state of 4U 1630$+$47 is recently reported which seems to be originated from a disc-wind regulated accretion scenario \cite[]{Ratheesh-etal2023, Kushwaha-etal2023}. However, no such disc-wind signature is observed from the spectro-polarimetric study of LMC X$-3$.

We find increasing PDs in higher energy band as compared to the expected nature from the existing models \cite[]{Schnittman-etal2009, Taverna-etal2020}. In reality, the pure electron scattering with high binary inclination of $\gtrsim 70^{\circ}$ can lead to such high PDs \cite[]{Chandrasekhar1960,Ratheesh-etal2023}. However, the complete ionization of matter assumed in pure electron-scattering theory is unlikely to achieve in accretion disc. Alternatively, in a partially-ionized disc atmosphere along with a weak corona, the absorption effect significantly enhances PD up to $\sim 4\%$ for a lower inclination $\lesssim 71^{\circ}$ and hence, the energy dependent polarization can be explained in a self-similar way \cite[]{Ratheesh-etal2023}. With this, we infer that the observed PD $\sim 3.04\%$ of LMC X$-3$ for a disc inclination $\sim 69^{\circ}$ is possibly resulted due to the direct and reflected emissions in a partially-ionized disc atmosphere. Further, the broad-band spectral modelling (see \S3.4) indicates that LMC X$-3$ seems to be surrounded by a slim disc structure that renders the polarized emission \cite[]{Ratheesh-etal2023}.

Finally, we constrain the black hole spin ($a_{*}$) in LMC X$-3$ using broad-band spectral modelling. We find that different disc models yield $a_{*}$ as $0.295_{-0.021}^{+0.008}$ and $0.273_{-0.012}^{+0.011}$ at $90\%$ confidence, respectively. These results are in agreement with the previous estimates \citep{Steiner-etal2010, Bhuvana-etal2021}. Moreover, we find that LMC X$-3$ accretes at sub-Eddington rate ($4.27_{-0.13}^{+0.12} \times 10^{18}$ g s$^{-1}$) for source mass $M_{\rm BH} = 6.98 ~ {\rm M}_\odot$.

In summary, in this endeavor, we report the first detection of polarized emissions in a thermally dominated persistent extra-galactic BH-XRB, namely LMC X$-3$.

\section*{Acknowledgments}

Authors thank the anonymous reviewer for constructive comments and useful suggestions	that help to improve the quality of the manuscript. SD acknowledges Science and Engineering Research Board (SERB), India for support under grant MTR/2020/000331. AK and AN thank GH, SAG; DD, PDMSA, and Director, URSC for encouragement and continuous support to carry out this research. This publication uses the data from the {\it IXPE}, {\it NICER} and {\it NuSTAR} missions, archived at the HEASARC data centre. 

\section*{Data Availability}

Data used for this publication are currently available at the HEASARC browse website (\url{https://heasarc.gsfc.nasa.gov/db-perl/W3Browse/w3browse.pl}).

\input{ms.bbl}

\end{document}